\documentclass[%
reprint,
amsmath,amssymb,
aps,
]{revtex4-2}

\usepackage{graphicx}
\usepackage{dcolumn}
\usepackage{bm}
\usepackage{xcolor}
\usepackage{comment}

\usepackage[normalem]{ulem}

\begin{document}
	
\preprint{APS/123-QED}
	
\title{Non-linear Structure Formation in Planck+DESI Favoured Interacting Dark Energy Cosmologies}

\author{Yuejia Zhai}
\email{yzhai13@sheffield.ac.uk}
\affiliation{School of Mathematical and Physical Sciences, University of Sheffield, Hounsfield Road, Sheffield S3 7RH, United Kingdom}
	
\author{Carsten van de Bruck}
\email{c.vandebruck@sheffield.ac.uk}
\affiliation{School of Mathematical and Physical Sciences, University of Sheffield, Hounsfield Road, Sheffield S3 7RH, United Kingdom}
	
\author{Eleonora Di Valentino}
\email{e.divalentino@sheffield.ac.uk}
\affiliation{School of Mathematical and Physical Sciences, University of Sheffield, Hounsfield Road, Sheffield S3 7RH, United Kingdom}

\author{Baojiu Li}
\email{baojiu.li@durham.ac.uk}
\affiliation{Institute for Computational Cosmology, Department of Physics, Durham University, South Road, Durham DH1 3LE, United Kingdom}

\author{Rafael C. Nunes}
\email{rafadcnunes@gmail.com}
\affiliation{Instituto de F\'{i}sica, Universidade Federal do Rio Grande do Sul, 91501-970 Porto Alegre RS, Brazil}
\affiliation{Divis\~{a}o de Astrof\'{i}sica, Instituto Nacional de Pesquisas Espaciais, Avenida dos Astronautas 1758, S\~{a}o Jos\'{e} dos Campos, 12227-010, S\~{a}o Paulo, Brazil}
	
	\date{\today}
	
\begin{abstract}    
Following our previous work constraining interacting dark energy (IDE) models, which showed their potential to alleviate the Hubble tension, in this work we investigate the non-linear effects of the IDE scenario favoured by CMB and DESI observations. The implications of IDE for the $S_8$ tension remain unclear, since current weak-lensing and large-scale-structure analyses either exclude highly non-linear scales or model the non-linear regime using prescriptions calibrated within $\Lambda$CDM. We address this issue by implementing a fully self-consistent IDE pipeline.
We perform N-body simulations of the IDE model with a transfer rate $Q=\xi {\cal H}\rho_x$ using a modified implementation of \texttt{RAMSES}. Since the dark matter Euler equation remains unchanged with respect to $\Lambda$CDM, the interaction can be incorporated through the modified background evolution and an effective time-dependent dark matter particle mass. We find scale-dependent deviations in the quasi-linear and non-linear regimes of the matter power spectrum, together with modifications to the density-field morphology and halo abundance. Our results show that the impact of IDE on quasi-linear and non-linear structure formation cannot be captured by standard $\Lambda$CDM-calibrated prescriptions, highlighting the importance of model-consistent non-linear modelling for future weak-lensing and large-scale-structure constraints on interacting dark energy cosmologies.
\end{abstract}
	
\maketitle
	
	
\section{\label{sec:Intro}Introduction}
	
$\Lambda$CDM has been remarkably successful in explaining a wide range of cosmological observations, including supernovae~\cite{SupernovaSearchTeam:1998fmf,SupernovaCosmologyProject:1998vns}, the cosmic microwave background (CMB)~\cite{WMAP:2012nax,Planck:2018vyg, Planck:2018nkj,ACT:2020gnv,ACT:2025fju,SPT-3G:2025bzu}, and galaxy clustering in the local Universe~\cite{BOSS:2016wmc,DESI:2025fii,DESI:2024mwx}. Despite this success, discrepancies have emerged among cosmological parameters inferred from different probes, most notably the Hubble tension~\cite{Verde:2019ivm,DiValentino:2020zio,DiValentino:2021izs,Perivolaropoulos:2021jda}, and the debated $S_8$ tension, with DES Year 6 results showing a $2.4\sigma$--$2.7\sigma$ tension in $\sigma_8$, after KiDS Legacy reported results consistent within $1\sigma$ with the \textit{Planck} Legacy constraints~\cite{Heymans:2020gsg,Wright:2025xka,DES:2026fyc,Pantos:2026koc}. While these discrepancies may still be affected by residual systematics, their persistence across independent analyses has motivated extensive investigation of physics beyond the standard cosmological model~\cite{Abdalla:2022yfr,CosmoVerseNetwork:2025alb,Pantos:2026koc}.

In the fiducial cosmological model, the Universe is composed of baryonic matter, radiation, dark matter, and dark energy. In the absence of non-gravitational interactions between the different components, each energy-momentum tensor is separately conserved,
$\nabla_\mu T^{\mu}_{(i)\nu}=0$,
where $i$ labels the individual species: $x$ for dark energy and $c$ for cold dark matter. However, while the total energy-momentum tensor must be conserved, separate conservation of the dark matter and dark energy sectors is not required by fundamental principles. This opens the possibility of an interaction within the dark sector.

The possibility of interactions within the dark sector has been extensively studied over the past three decades~\cite{Wetterich:1994bg,Amendola:1999er,Zimdahl:2001ar,Farrar:2003uw,Brookfield:2007au,VanDeBruck:2017mua}. A theoretically motivated realization arises in coupled quintessence models, where an interaction between dark matter and a scalar-field dark energy component can emerge, for instance, through conformal or disformal transformations of scalar-tensor theories~\cite{Amendola:1999qq,Amendola:1999er,Zumalacarregui:2010wj,vandeBruck:2015ida}. A complementary phenomenological approach introduces a source term directly into the background conservation equations, allowing energy exchange between the dark components~\cite{Zimdahl:2001ar}. In recent years, renewed observational interest in interacting dark energy (IDE) models has been driven by their potential impact on cosmological tensions, in particular the Hubble tension~\cite{Salvatelli:2013wra,Kumar:2016zpg,Caprini:2016qxs,Murgia:2016ccp,Zheng:2017asg,Kumar:2017dnp,DiValentino:2017iww,Kumar:2021eev,Gao:2021xnk,Pan:2023mie,Benisty:2024lmj,Yang:2020uga,Forconi:2023hsj,Pourtsidou:2016ico,DiValentino:2020vnx,DiValentino:2020leo,Nunes:2021zzi,Yang:2018uae,Zhang:2018mlj,vonMarttens:2019ixw,Lucca:2020zjb,Xiao:2021nmk,Wang:2024vmw,Gao:2022ahg,Zhai:2023yny,Joseph:2022khn,Bernui:2023byc,Becker:2020hzj,Hoerning:2023hks,Giare:2024ytc,Mukhopadhyay:2020bml,Escamilla:2023shf,vanderWesthuizen:2023hcl,Silva:2024ift,DiValentino:2019ffd,Zhao:2022ycr,Li:2024qso,Pooya:2024wsq,Halder:2024uao,Castello:2023zjr,Yao:2023jau,Mishra:2023ueo,Nunes:2016dlj,Silva:2025hxw,Yang:2025uyv,vanderWesthuizen:2025rip,Zhang:2025dwu,Li:2025muv,Li:2025owk,Li:2026xaz}.

In this work we focus on the phenomenological IDE model characterised by a dark-sector interaction rate $Q$ proportional to the dark energy density $\rho_x$, $Q=\xi {\cal H}\rho_x$, where ${\cal H}$ is the conformal Hubble parameter and $\xi$ is a dimensionless coupling parameter. This class of models has attracted particular attention because it can shift the CMB-inferred value of $H_0$ towards larger values and thereby reduce the Hubble tension~\cite{LopezHonorez:2010esq,DiValentino:2019jae,Zhai:2023yny}. More generally, CMB and BAO constraints on phenomenological interactions of the form $Q=\xi{\cal H}\rho_c$, $Q=\xi{\cal H}\rho_x$, or $Q=\xi{\cal H}(c_1\rho_c+c_2\rho_x)$ have shown that interactions in the dark sector can significantly affect the inferred expansion history and late-time cosmological parameters~\cite{Zhai:2023yny}.
However, the effects of these models on non-linear structure formation have been little explored in the literature, which makes it crucial to analyse current cosmological tracers, such as weak lensing, galaxy clustering, and other probes, especially for establishing robust constraints on $S_8$. Some recent semi-analytical approaches in this context were presented in~\cite{Silva:2024ift,Silva:2025bnn}, for example.

Understanding the effects of IDE cosmology on non-linear scales is therefore an important issue, because IDE models that reduce the Hubble tension often allow, or even prefer, relatively large values of $S_8$, but these constraints are typically weak and strongly dependent on the treatment of non-linear structure formation. In current weak-lensing and large-scale-structure analyses, the most non-linear scales are often removed or modelled using fitting functions and prescriptions calibrated within $\Lambda$CDM. This is not fully self-consistent for an IDE cosmology, since the interaction modifies the background evolution, the growth history, and potentially the mapping between linear and non-linear clustering. As a result, it remains unclear whether the apparent behaviour of IDE with respect to the $S_8$ tension reflects a genuine physical prediction of the model or limitations of the non-linear modelling adopted in current analyses.

Models allowing for dark-sector interactions have been explored through N-body simulations in~\cite{Cui:2012is,Baldi:2010pq,Baldi:2011qi,Zhang:2018mlj,Zhao:2022ycr}. Similarly, here we perform robust gravity-only N-body simulations of the IDE model with coupling $Q=\xi {\cal H}\rho_x$, using a model-consistent implementation of the modified background and dark matter evolution. This provides a step towards a more accurate treatment of the full matter power spectrum, which is essential for future observational analyses that require $P(k,z)$ as a primary physical input in this class of interacting dark energy scenarios.

In Sec.~\ref{sec:Theory}, we review the interacting dark energy theory relevant to this work. In Sec.~\ref{sec:Methods}, we describe the implementation of the model in the numerical pipeline, including the Boltzmann solver and the N-body simulation code. In Sec.~\ref{sec:Results}, we present the simulation results and focus on the impact of the interaction on non-linear structure formation. Finally, in Sec.~\ref{final}, we conclude with a summary of our results and future perspectives.

\section{Theory}
\label{sec:Theory}

In the conformal Newtonian gauge, we work with a Friedmann-Robertson-Walker metric of the form
\begin{equation}
	ds^2 = a^2 \left[-(1+2\Phi)d\tau^2+(1-2\Psi)\delta_{ij}dx^i dx^j \right]~.
\end{equation}

While the total energy-momentum tensor remains conserved, the interacting dark energy model introduces an energy-momentum transfer between dark matter and dark energy as follows
\begin{equation}
\label{IDE_eq_cont}
    \nabla_\mu T^{\mu}_{c~\nu}=Q_\nu, 
    \qquad 
    \nabla_\mu T^{\mu}_{x~\nu}=-Q_\nu,
\end{equation}
where $Q_\nu$ controls the transfer of energy and momentum between dark matter and dark energy. As mentioned in the introduction, the specific IDE model studied in this paper is characterised by
\begin{equation}
Q = \xi \mathcal{H}\rho_x~,
\end{equation}
where $\mathcal{H}$ is the conformal Hubble parameter, $\rho_x$ is the dark energy density, and $\xi$ is a dimensionless coupling parameter. For $\xi<0$, energy is transferred from dark matter to dark energy, while for $\xi>0$ the energy flow is reversed. This model has been extensively studied in the context of cosmological tensions and has been shown to shift the CMB-inferred value of $H_0$ towards larger values (e.g.~\cite{DiValentino:2019jae,Zhai:2023yny}).

The perturbation equations for this interacting scenario follow the implementation developed in previous works~\cite{Valiviita:2008iv,DiValentino:2019jae,Zhai:2023yny}. The equations for the DM component are given in Fourier space by
\begin{align}
	\dot\delta_c &= -\left( \theta_c - 3\dot\Phi \right) + \frac{Q}{\rho_c}\left(\frac{\delta Q}{Q}  - \delta_c + \Psi \right)~, \\
	\dot\theta_c &= -{\cal H}\theta_c + k^2 \Psi~,
\end{align}
where $\delta_c$ is the dark matter density contrast and $\theta_c$ is the divergence of the dark matter velocity field. We point out that, for the class of models considered here, where the energy-momentum transfer is parallel to the dark matter four-velocity, the Euler equation for dark matter remains unmodified with respect to $\Lambda$CDM. As a consequence, no additional fifth force arises in the dark matter dynamics. This property plays a crucial role in the N-body implementation adopted in this work, since the non-linear evolution can be treated using standard Newtonian particle dynamics with a modified background expansion and effective dark matter density evolution, which we implement in the \texttt{RAMSES} code.

The interaction affects the growth of perturbations both through the modified expansion history and through the evolution of the dark matter density. In principle, dark energy perturbations also contribute to the Poisson equation,
\begin{equation}
	k^2\Psi = -4\pi G a^2 \left( \rho_c\delta_c + \rho_x\delta_x \right)~.
\end{equation}
To quantify the importance of dark energy clustering, we define
\begin{equation}
	\beta(k,z)\equiv
	\frac{\rho_x}{\rho_c}
	\frac{\delta_x}{\delta_c}~.
\end{equation}

For the IDE parameter space explored in this work, we find $\beta(k,z)\ll1$ on the scales relevant for the simulations. This allows us to safely neglect dark energy clustering in the Poisson equation and treat the interaction effectively through the modified background quantities and dark matter evolution.

In the context of the N-body simulations performed in this work, the energy transfer between dark matter and dark energy is interpreted as an effective evolution of the dark matter particle mass. Writing the dark matter density within a sufficiently large simulation volume $V$ as
\begin{equation}
\rho_c=\frac{Nm_c}{a^3 V}~,
\end{equation}
where $N$ is the number of DM particles within $V$ and $m_c$ is an effective time-dependent DM particle mass, the continuity equation implies
\begin{equation}
Q=\frac{\dot m_c}{m_c}\rho_c~.
\end{equation}
This leads to
\begin{equation}
\frac{m_c(a)}{m_c(a_0)}
=
\exp\left[
\xi
\int_{a_0}^{a}
\frac{\rho_x}{\rho_c}
d\ln a
\right]~.
\end{equation}
The mass-rescaling quantity $m_c(a)/m_c(a_0)$, together with the modified expansion history $H(a)$, fully encapsulates the effects of the interaction between dark matter and dark energy. By absorbing the interaction into a time-dependent particle mass and a modified expansion rate, the dark matter density averaged over a sufficiently large volume satisfies the energy-conservation equation of the cosmological background. Thus, in the context of N-body simulations, the model considered here has time-dependent dark matter masses but does not predict a fifth force mediated by dark energy, as is the case in other models, such as coupled quintessence~\cite{Amendola:2003wa,Maccio:2003yk,Baldi:2008ay}.

\section{\label{sec:Methods} Methodology}
    
At early times, the Universe is matter dominated and the interaction has a negligible impact on the growth of cosmic structures. In particular, for the class of IDE models considered in this work, the contribution of dark energy perturbations to the gravitational potential remains subdominant on the scales relevant to the simulations. This allows us to adopt a simplified, yet physically well-motivated, implementation of the non-linear evolution.

Since the dark matter Euler equation remains unchanged with respect to $\Lambda$CDM, the particle dynamics can be evolved using standard Newtonian N-body methods. The effects of the interaction are incorporated through the IDE initial conditions, the modified background expansion history, and the evolving mass of the dark matter particles. The resulting approach provides a model-consistent treatment of non-linear structure formation in the regime where dark energy clustering can be neglected.

\subsection{Initial Conditions}
	
The initial conditions for our N-body simulations are generated using the publicly available \texttt{2LPTic} code~\cite{Crocce:2006ve}, which displaces particles from a uniform grid according to second-order Lagrangian perturbation theory (2LPT). The 2LPT approach provides more accurate initial conditions than the Zel'dovich approximation, as it captures the onset of non-linear gravitational collapse through second-order corrections to the displacement field, thereby reducing transient effects from the initialisation at moderate redshifts.

The displacement field is constructed by drawing Fourier modes from the input power spectrum with a fixed random seed, guaranteeing that both the IDE and $\Lambda$CDM initial conditions share identical initial phases and differ only in the amplitude and shape of the power spectrum as predicted by their respective linear theories. This choice allows a direct comparison between the non-linear evolution in the two cosmologies while minimising cosmic variance between the simulations. The initial conditions for each model are generated at the redshift $z_{\rm ini} = 49$ by setting the linear growth factor $D=1$ and normalising to $\sigma(R=8h^{-1}{\rm Mpc},z=49)$.

The input power spectrum is the linear matter power spectrum computed at the initial redshift $z_{\rm ini} = 49$ using the modified version of the Boltzmann solver \texttt{CLASS}~\citep{Blas:2011} from our previous works~\cite{DiValentino:2019jae,Zhai:2023yny}. The cosmological parameters adopted for these initial conditions correspond to the best-fit values obtained from the Bayesian analysis of the \textit{Planck} 2018 TTTEEE+lensing likelihood~\cite{Planck:2019nip,Planck:2018vyg,Planck:2018nkj,Planck:2018lbu}, DESI DR2 BAO~\cite{DESI:2025zgx}, and a Gaussian prior on the Hubble constant from the SH0ES collaboration, $H_0 = (73.04 \pm 1.04)$ km/s/Mpc~\cite{Riess:2021jrx}.

For the $\Lambda$CDM reference simulation, we instead use the standard \texttt{CLASS} implementation~\citep{Blas:2011} and adopt the best-fit values from the Bayesian analysis of \textit{Planck} 2018 TTTEEE+lensing~\cite{Planck:2019nip,Planck:2018vyg,Planck:2018nkj,Planck:2018lbu}. Both power spectra are evaluated at $z_{\rm ini} = 49$, where the interaction has only a limited impact on the growth of structure and the density fields remain safely in the quasi-linear regime.
	
	\begin{table}
\caption{\label{tab:table1} Cosmological parameters for the $\Lambda$CDM model from P18+lensing and for the IDE model from P18+lensing + DESI DR2 BAO + SH0ES. The $\sigma_8$ values reported in the table correspond to the best-fit values from the above datasets and are therefore not free input parameters. While reported for completeness, $\sigma_8$ is not directly used as an input parameter in the N-body simulations.}
		\begin{ruledtabular}
			\begin{tabular}{lcc}
				Model & IDE&$\Lambda$CDM\\
				\hline
				\textbf{[setup]} & &\\
				starting redshift & 49&49\\
				$\ell_{min}$ (coarse level) & 8 &8\\
				$\ell_{max}$ (fine level) & 16& 14 to 16\\
				\textbf{[cosmology]} & &\\
				$\Omega_m$                 & 0.2191&0.3158\\
				$\Omega_\Lambda$                 & 0.7809&0.6842\\
                $\ln 10^{10} A_s$& $3.0395$ & $3.0448$\\
                $n_s$& $0.9701$ & $0.9661$ \\
				$\xi$ & -0.27 & -\\
				$H_0$(km $s^{-1}Mpc^{-1}$)                      & 70.63&67.32\\
				$\sigma_8$ &1.0674&0.8120\\
			\end{tabular}
		\end{ruledtabular}
	\end{table}

	\begin{table}[b]
		\caption{\label{tab:table2}%
For each of the IDE and $\Lambda$CDM models, we performed two simulations using the cosmological parameters listed in Table~\ref{tab:table1}. The corresponding simulation settings are summarised below.
}
		\begin{ruledtabular}
			\begin{tabular}{cccc}
				Model&Box size $L_{box}$& Particle number $N_p^3$& Setting \\
				& ($\mathrm{Mpc}/h$)& &\\
				\hline
				IDE&$256$&$256^3$& A\\
				&$512$&$512^3$&B\\
				$\Lambda$CDM&$256$&$256^3$&A\\
				&$512$&$512^3$&B\\
			\end{tabular}
		\end{ruledtabular}
	\end{table}

\subsection{Simulations}

We performed N-body simulations using the adaptive mesh refinement (AMR) code \texttt{RAMSES}~\cite{Teyssier:2001cp}, modified to incorporate the IDE cosmology as described in Section~\ref{sec:Theory}. N-body simulations track the gravitational evolution of $N_p^3$ dark matter particles inside a periodic comoving volume of side length $L_{\rm box}$.

Table~\ref{tab:table2} summarizes the numerical configurations adopted for both the IDE and $\Lambda$CDM scenarios.

We performed simulations using two configurations. A preliminary run with $L_{\mathrm{box}}=256\,\mathrm{Mpc}/h$ and $256^3$ particles (hereafter setting~A) was used to inspect and validate the implementation against the linear predictions from our modified \texttt{CLASS} code. The production run adopts $L_{\mathrm{box}}=512\,\mathrm{Mpc}/h$ with $512^3$ particles (hereafter setting~B). The largest scale accessible in the matter power spectrum is set by the fundamental mode, $k_f = 2\pi/L_{\rm box}$. In practice, the spectrum is considered reliable only for $k \gtrsim k_{\rm min} \simeq 2k_f$~\cite{Schneider:2015yka}, giving $k_{\rm min} \approx 0.049\,h\,\mathrm{Mpc}^{-1}$ and $0.025\,h\,\mathrm{Mpc}^{-1}$ for the small and large boxes, respectively. On small scales, the resolution is limited by the particle Nyquist frequency, $k_{\rm max} \simeq k_{\rm Ny} = \pi N_p/L_{\rm box} \approx 3.1\,h\,\mathrm{Mpc}^{-1}$ for both configurations.

The gravitational evolution is governed by standard Newtonian dynamics, since the IDE model considered here does not modify the dark matter Euler equation and therefore does not introduce a fifth force. The interaction is incorporated into the simulations through the following ingredients:
\begin{itemize}
\item the modified initial matter power spectrum, which encodes the IDE linear growth history;
\item the modified background expansion history, $H(a)$;
\item the effective time-dependent dark matter particle mass, $m_c(a)$.
\end{itemize}

\begin{figure*}
        \centering
        \includegraphics[width=\columnwidth]{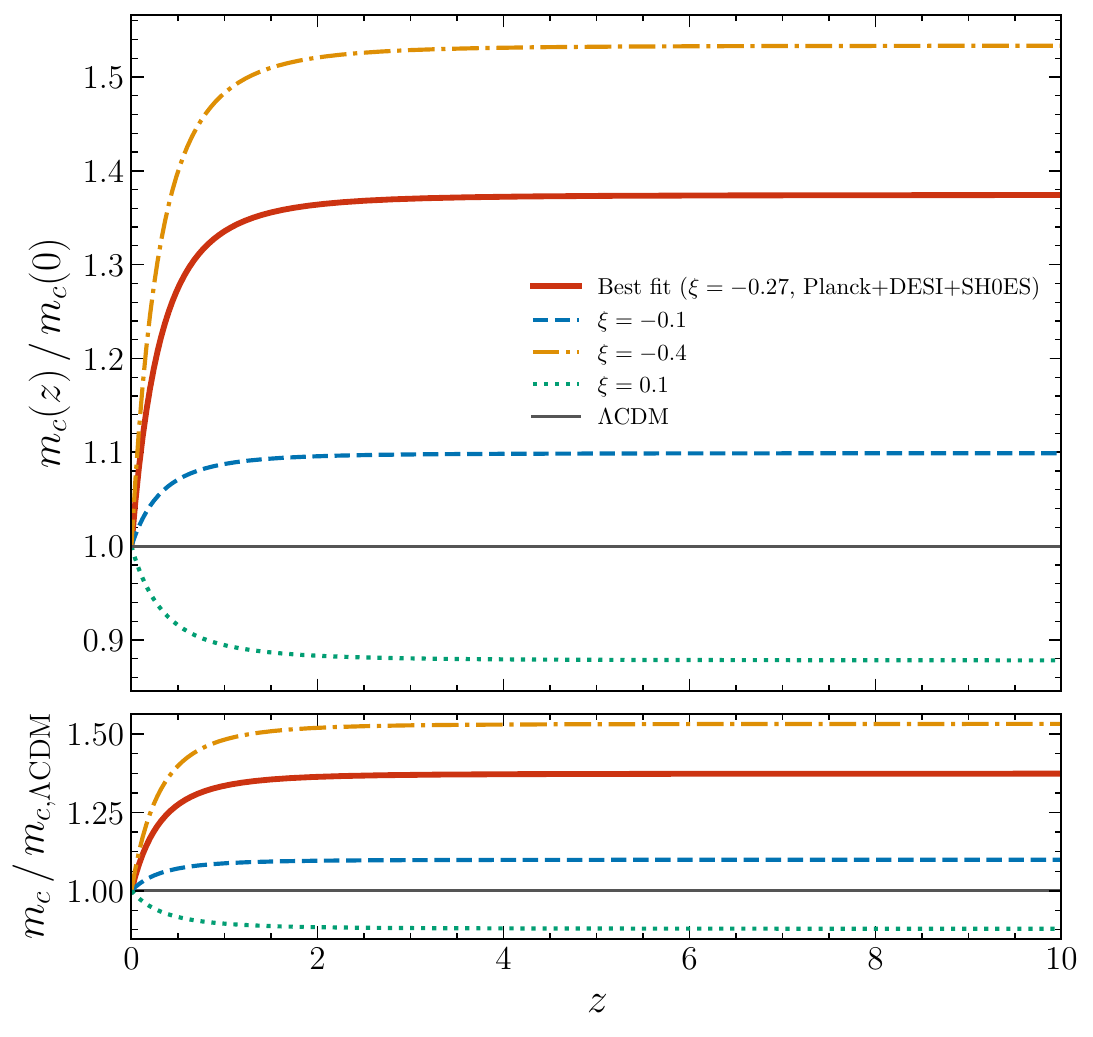}
        \includegraphics[width=\columnwidth]{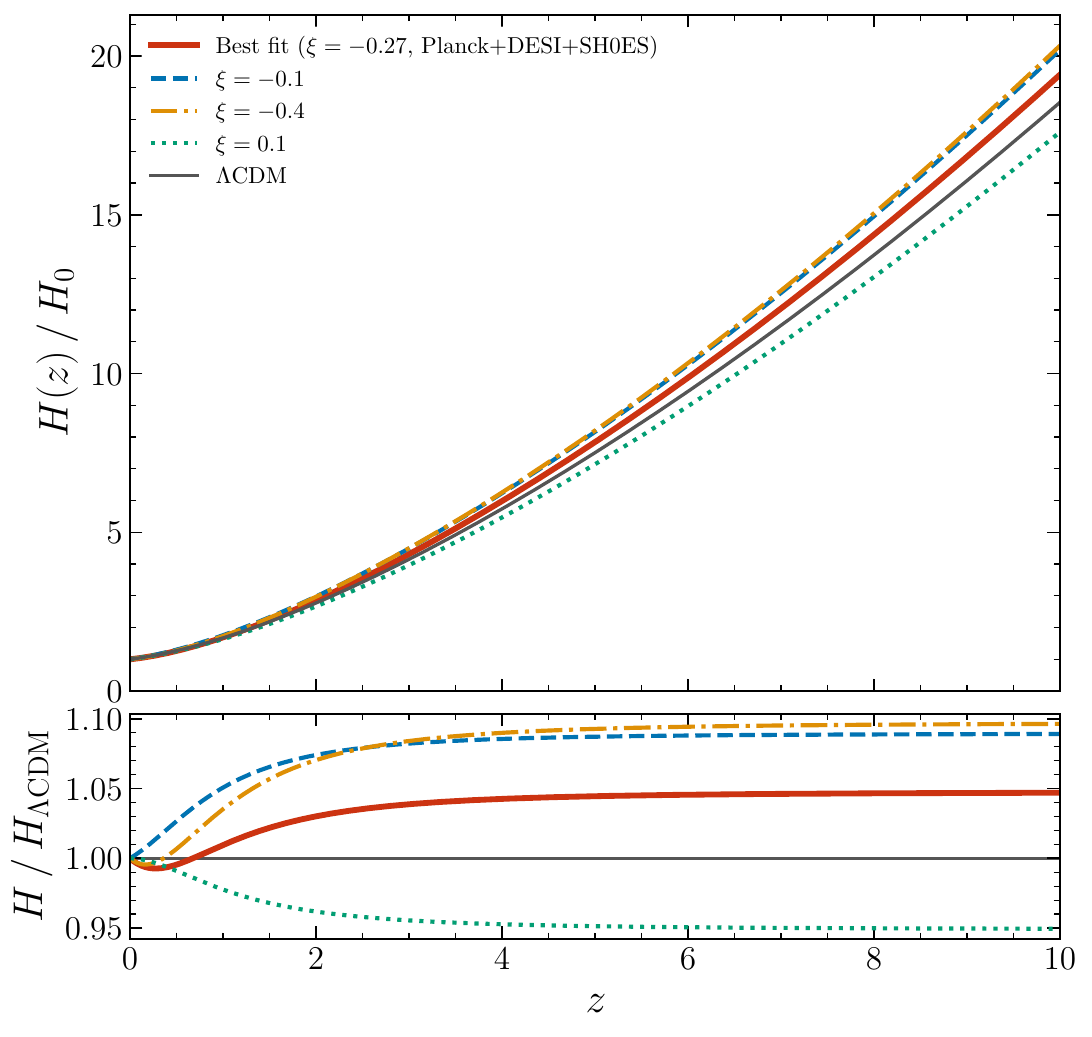}
        \caption{Evolution of the dark matter particle mass and Hubble parameter in interacting dark energy models. The left panel shows the ratio of the CDM particle mass, $m_c(z)$, to its present-day value, $m_c(0)$, as a function of redshift for various coupling strengths $\xi$. For the best-fit value $\xi=-0.27$ from \textit{Planck}+DESI+SH0ES (solid red), the DM particle mass at high redshift is $\sim37\%$ larger than at $z=0$, reflecting energy transfer from dark matter to dark energy. The right panel shows the evolution of the Hubble parameter, $H(z)/H_0$. The lower subpanels show the ratio with respect to $\Lambda$CDM. Negative (positive) values of $\xi$ correspond to energy flow from DM to DE (DE to DM).}
		\label{fig:bg_evolution}
\end{figure*}

The modified background evolution compared to the fiducial $\Lambda$CDM case is shown in Figure~\ref{fig:bg_evolution}. The evolution of $m_c(z)/m_c(z=0)$ and $H(z)/H_0$ for four cosmological settings, including a $\Lambda$CDM case for reference, is presented. The ``Best fit'' refers to the IDE scenario for which we adopt the best-fit values from the Bayesian inference results of P18+lensing+DESI with a Gaussian prior on $H_0$.

The expansion history is provided to \texttt{RAMSES} through a tabulated file specifying $m_c(a)/m_c(a_0)$, $H(a)$, and the scale factor $a$, ensuring that the simulation captures the altered Hubble drag experienced by dark matter particles in the IDE cosmology. The particle masses are rescaled consistently using the tabulated ratio $m_c(a)/m_c(a_0)$ each time the Poisson equation is solved, accounting for the modified dark matter density evolution induced by the interaction.

From the simulation outputs at $z=0$, we extract the dark matter particle distribution to identify collapsed structures. Halos are identified using the HOP (Hierarchical Overdensity-based Particle) algorithm~\cite{Eisenstein:1997sm} from the publicly available \texttt{yt} Astrophysical Analysis Extension~\citep{Turk:2010ah,yt.astro.analysis}. HOP identifies halos as density peaks and assigns particles to halos by following local density gradients. Since this procedure yields halo masses that are broadly comparable to spherical-overdensity (SO) definitions, we compare our results to the SO-calibrated fitting function of~\cite{Despali:2015yla}, which adopts a Sheth-Tormen-like form~\cite{Sheth:1999mn},
\begin{equation}
    f(\sigma) = 2 A \left(1 + \frac{1}{(a\nu)^p}\right)
    \left(\frac{a\,\nu}{2\,\pi}\right)^{1/2}
    e^{-a\nu/2} ,
    \label{eq:despali_fsigma}
\end{equation}
where $\nu \equiv \delta_c^2/\sigma^2$ is the peak height and $\delta_c\simeq1.686$ is the linear collapse threshold. The variance of the linear density field, smoothed on a scale $R$, is
\begin{equation}
    \sigma^2(R) = \frac{1}{2\pi^2}
    \int_0^\infty P(k)\,\hat{W}^2(kR)\,k^2\,dk ,
    \label{eq:sigma}
\end{equation}
where $P(k)$ is the linear matter power spectrum and $\hat{W}(kR)$ is the Fourier transform of the real-space top-hat window function. The free parameters $(A,a,p)$ are calibrated against N-body simulations for a given halo mass definition and redshift. We note, however, that this fitting function is calibrated on standard $\Lambda$CDM N-body simulations and should therefore be interpreted as a reference prediction rather than a model-consistent IDE calibration.

The differential halo mass function $dn/d\ln M$ is computed using narrow logarithmic mass bins of width $\Delta\ln M\approx 0.461$ narrow enough to keep the binning systematic well below $1\%$~\cite{Lukic:2007fc}, spanning the range from $M = 10^{12}$ to $10^{15.5}\,M_{\odot}h^{-1}$, as
\begin{equation}
    \frac{dn}{d\ln M}=\frac{\Delta N}{V\,\Delta\ln M}~,
\end{equation}
where $M$ is the mean halo mass of all halos within a given bin, $\Delta N$ is the number of halos in that bin, and $V$ is the simulation volume.

\section{\label{sec:Results}Results}

\begin{figure*}
	\centering
	\includegraphics[width=\columnwidth]{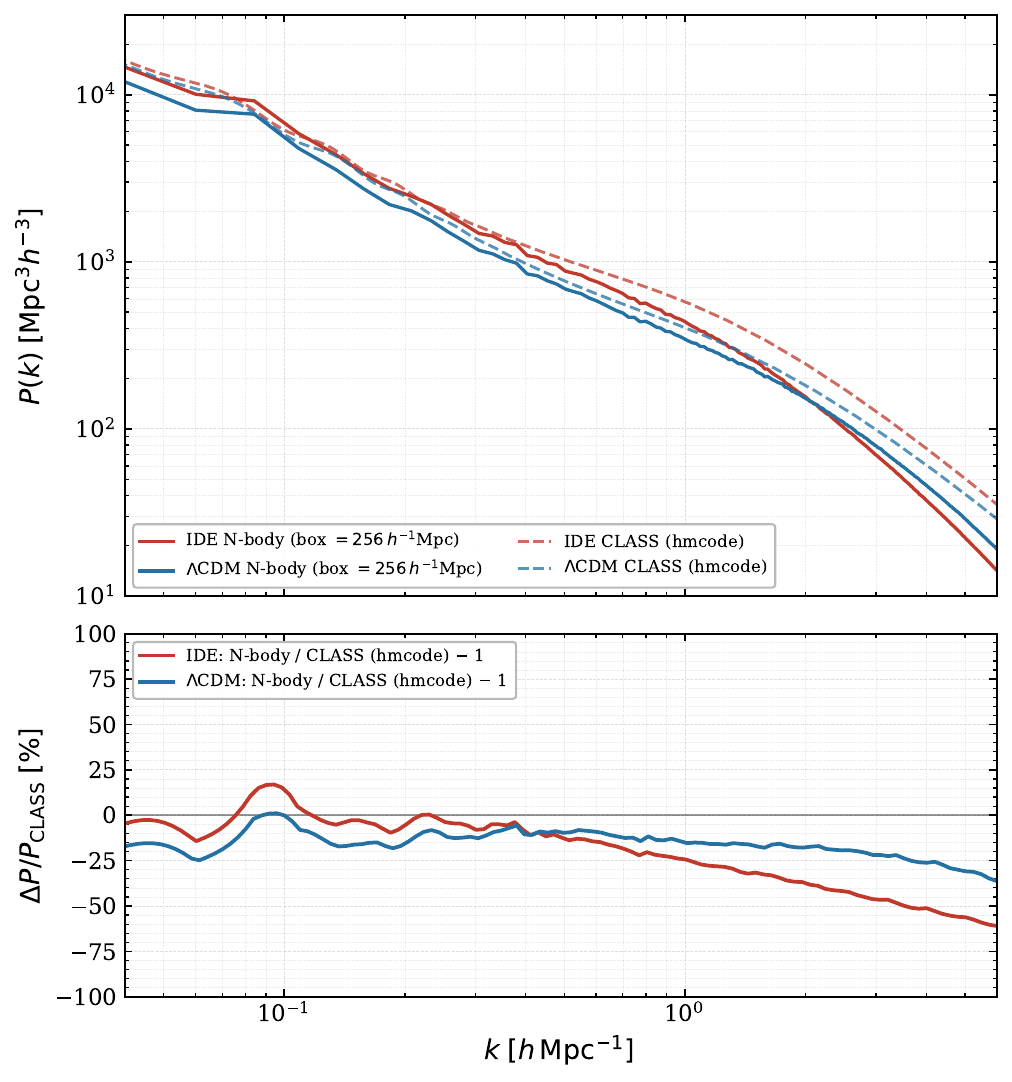}
	\includegraphics[width=\columnwidth]{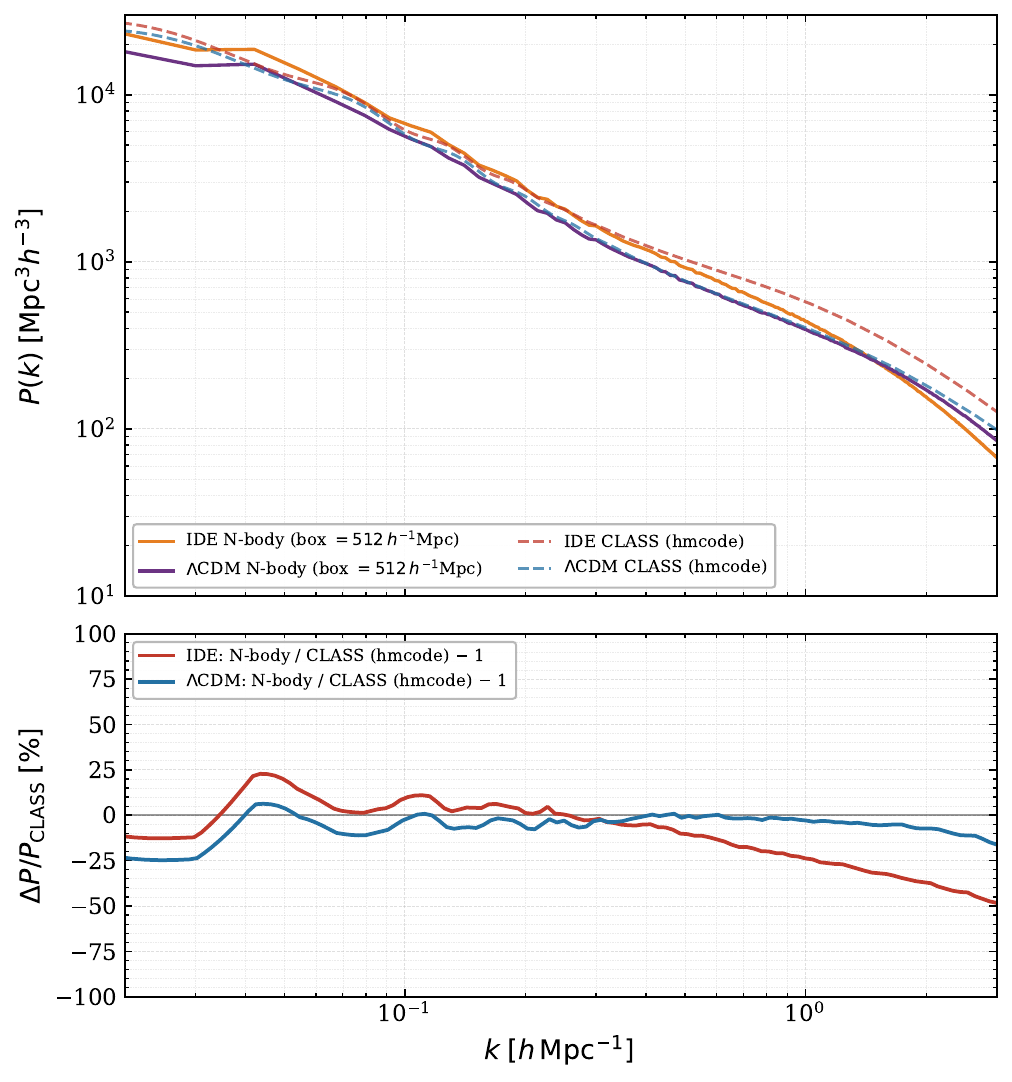}
	\caption{The left (right) panel shows the matter power spectra for the $\Lambda$CDM and IDE simulations in the $256\,\mathrm{Mpc}/h$ box ($512\,\mathrm{Mpc}/h$ box) at $z=0$. The solid (dashed) curves correspond to the matter power spectrum derived from the N-body simulations (computed with the \texttt{CLASS} code). The IDE coupling enhances clustering on large scales while suppressing it within collapsed haloes compared to the $\Lambda$CDM reference. The IDE linear matter power spectra (red dashed curves) have been rescaled by multiplying them by $(\sigma_{8,\Lambda\mathrm{CDM}}/\sigma_{8,\mathrm{IDE}})^2$ to bring the IDE and $\Lambda$CDM linear-theory predictions to the same normalisation, allowing a direct visual comparison of their scale dependence.}
	\label{fig:512}
\end{figure*}

Figure~\ref{fig:512} compares the matter power spectra obtained from the N-body simulations with the non-linear predictions from \texttt{CLASS} using the \texttt{HMcode} implementation. The results reveal several important aspects of non-linear structure formation in the IDE scenario considered in this work. The snapshots at $z=0$ allow us to simultaneously assess the physical impact of the dark-sector interaction and the numerical robustness of the simulations. On large scales, $k\simeq0.1\,h\,{\rm Mpc}^{-1}$, the N-body results closely follow the \texttt{CLASS} predictions, as expected in the quasi-linear regime. The small oscillations and amplitude deviations observed on these scales arise from sample variance, as the periodic box contains only a limited number of independent modes at low $k$.

Nevertheless, clear differences between IDE and $\Lambda$CDM are already visible. In particular, the IDE model exhibits enhanced clustering relative to $\Lambda$CDM on these scales. This behaviour is consistent with the modified linear growth history induced by the interaction, which leads to a larger amplitude of matter fluctuations. The quasi-linear regime, approximately $0.1\lesssim k \lesssim1\,h\,{\rm Mpc}^{-1}$, exhibits the most interesting behaviour. While the IDE model still shows enhanced clustering on relatively large scales, the non-linear evolution progressively suppresses this excess power. The lower panels of Fig.~\ref{fig:512} demonstrate that the fractional difference between the N-body results and the \texttt{CLASS+HMcode} predictions deviates significantly from zero as non-linear effects become important. This indicates that non-linear gravitational evolution modifies the mapping between the linear and non-linear matter power spectra in a non-trivial way in the IDE cosmology.

At smaller scales, $k\gtrsim1\,h\,{\rm Mpc}^{-1}$, the IDE simulations display a substantial suppression of power. This effect is significantly stronger than in the $\Lambda$CDM case and becomes particularly evident in the lower panels, where the IDE ratio decreases rapidly towards negative values at high $k$. Physically, this behaviour can be understood as a consequence of the effective evolution of the dark matter particle mass, $m_c(a)$, induced by the interaction. For the IDE model considered here, with $\xi<0$, energy is transferred from dark matter to dark energy, effectively reducing the amount of gravitating dark matter available at late times. As a result, although the modified expansion history enhances linear growth on large scales, the formation and virialisation of highly non-linear structures become less efficient.

An important consequence of this behaviour is that the IDE interaction enhances clustering on large scales while simultaneously suppressing power within massive haloes and inhibiting the accretion of smaller haloes onto more massive ones.

The comparison between the $256\,{\rm Mpc}/h$ (setting~A) and $512\,{\rm Mpc}/h$ (setting~B) simulations also demonstrates good numerical consistency. Both simulation volumes reproduce the same qualitative physical trends: enhanced clustering on large scales and suppression of power on small scales in the IDE cosmology. The setting~B results exhibit smoother behaviour and are more consistent with the \texttt{CLASS} and \texttt{HMcode} predictions at low $k$, as expected from the improved sampling of large-scale modes and reduced cosmic variance, while the small-scale suppression remains qualitatively stable between the two runs. This indicates that the main physical conclusions are robust against the numerical setup adopted in this work.

	\begin{figure*}
		\centering
		\includegraphics[width=\linewidth]{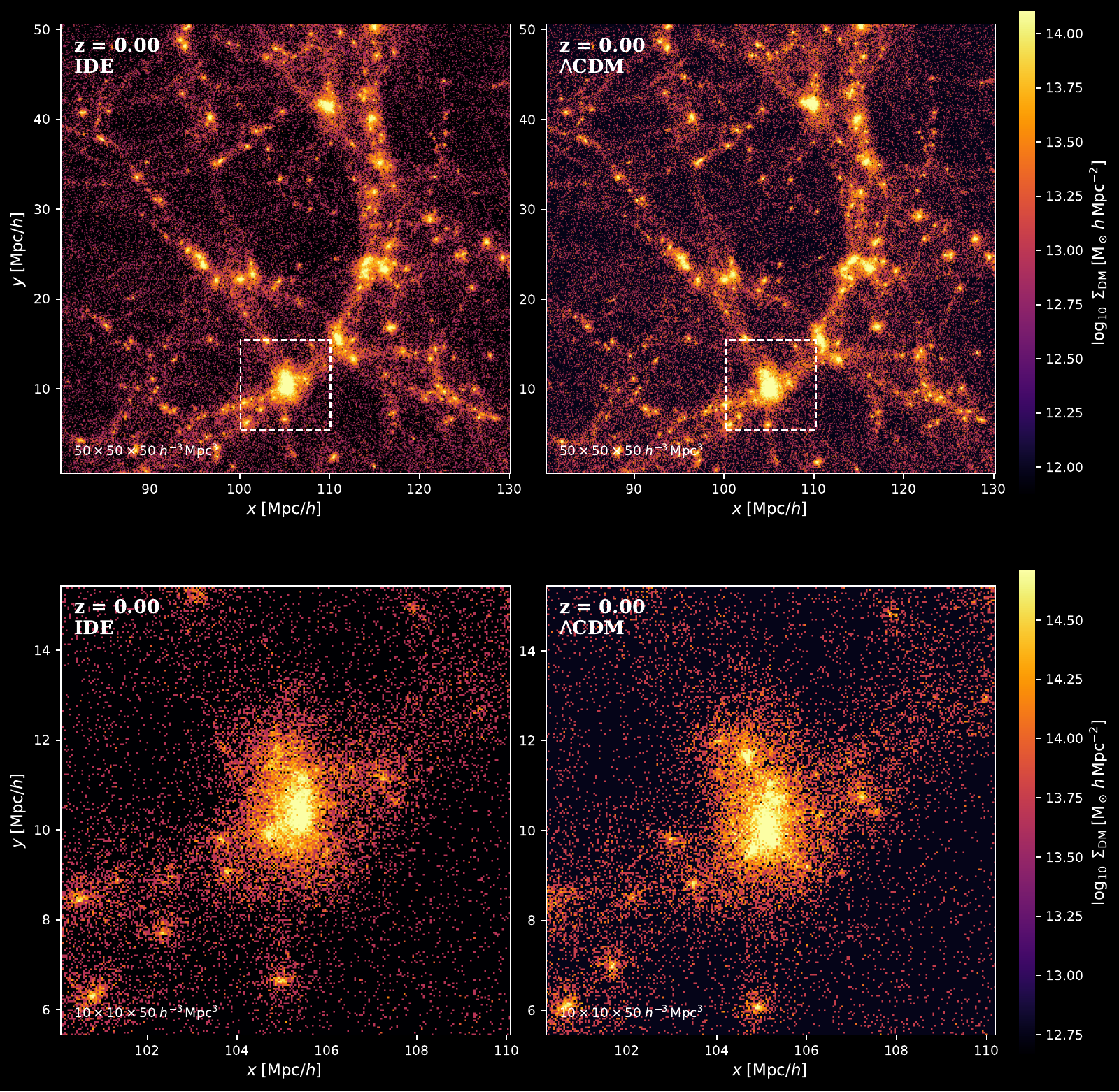}
		\caption{Projected dark matter density maps at $z=0$ for the IDE (left) and $\Lambda$CDM (right) cosmologies. The top row shows a $50\times50\times50\,h^{-3}\,{\rm Mpc}^3$ region centred on the most massive halo in each simulation, illustrating the large-scale filamentary structure. The bottom row shows a zoomed-in region of $10\times10\times50\,h^{-3}\,{\rm Mpc}^3$ centred on the same halo. The colour bar indicates the logarithmic value of the projected surface mass density, $\Sigma_{\rm DM}$. Both simulations use $256^3$ dark matter particles in a $(256\,h^{-1}\,{\rm Mpc})^3$ box with identical initial conditions.}
		\label{fig:halo_density}
	\end{figure*}
	
	\begin{figure}
		\centering
		\includegraphics[width=1.1\linewidth]{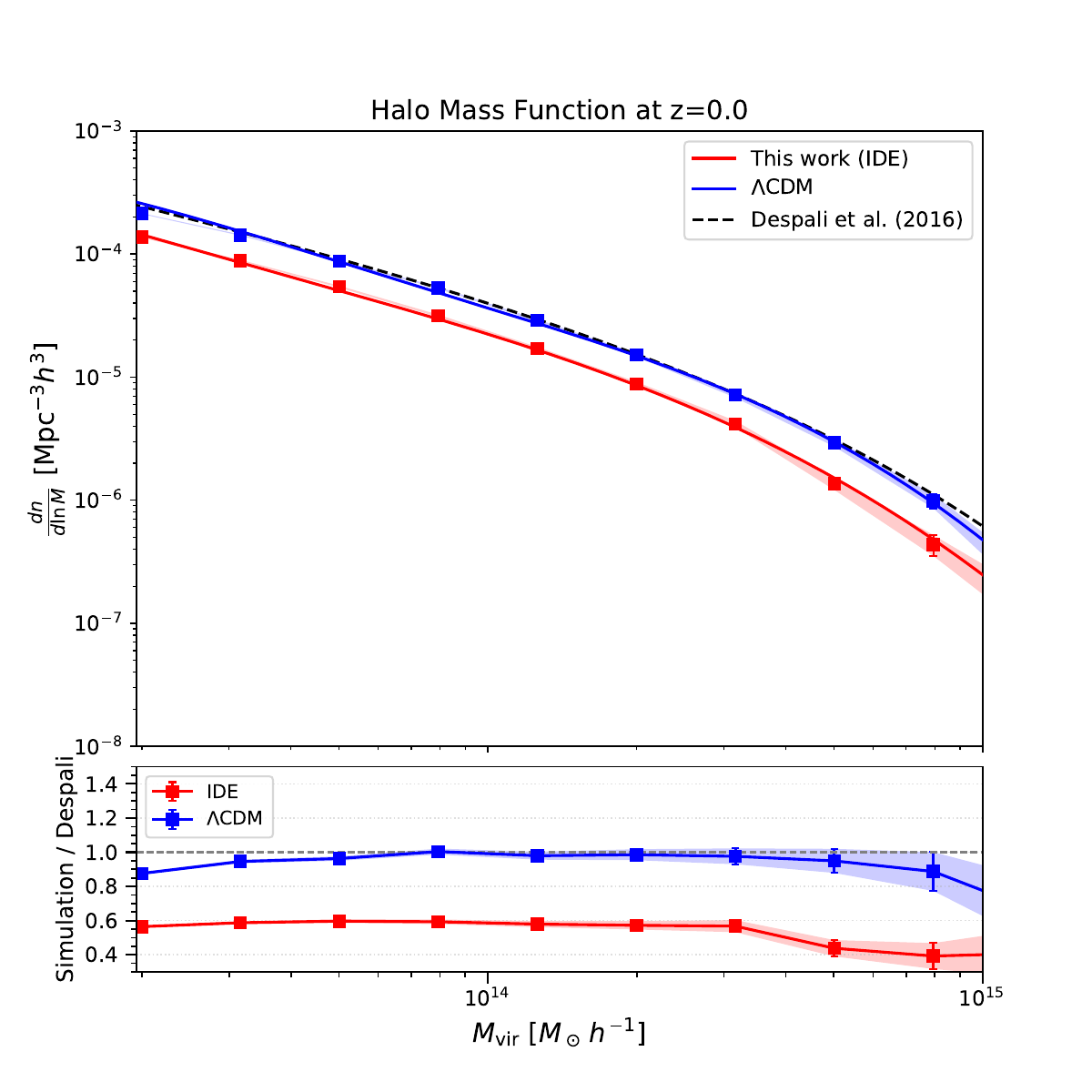}
		\caption{Halo mass functions (HMFs) at redshift $z=0$ from setting~B ($L_{\mathrm{box}}=512\,\mathrm{Mpc}/h$ with $512^3$ particles). The red solid curve shows the IDE simulation, the blue solid curve shows the $\Lambda$CDM reference, and the black dashed curve shows the fitting function from~\cite{Despali:2015yla} in Eq.~\ref{eq:despali_fsigma} for comparison. The error bars and shaded region show the Poisson uncertainty in each bin, $\sigma(dn/d\ln M)=\sqrt{\Delta N_{\rm bin}}/(V\cdot\Delta\ln M)$, where $\Delta N_{\rm bin}$ is the number of haloes in the selected bin.}
		\label{fig:hmf}
	\end{figure}

Figure~\ref{fig:halo_density} shows projected cold dark matter density maps at $z=0$ for the IDE (left panels) and $\Lambda$CDM (right panels) cosmologies. The top row displays a $50\times50\times50~h^{-3}{\rm Mpc}^3$ region centred on the most massive halo identified in each simulation, highlighting the large-scale filamentary structure of the cosmic web. Since both simulations were generated using identical initial phases, the overall morphology of the filament network remains remarkably similar in the two cosmologies. Filaments, nodes, and underdense regions appear at approximately the same locations, indicating that the IDE interaction does not strongly alter the geometry of the large-scale structure.
Despite these similarities, clear differences emerge in the non-linear regime. The $\Lambda$CDM simulation develops stronger overdensities at the intersections of filaments, producing brighter and more compact halo cores compared to the IDE case. In contrast, the IDE cosmology exhibits a visibly smoother matter distribution, with less concentrated central regions and slightly more diffuse filamentary environments around massive haloes.

The bottom panels present a zoom-in of a $10\times10\times50~h^{-3}{\rm Mpc}^3$ region centred on the same halo. At these smaller scales, the differences between the two cosmologies become more pronounced. The $\Lambda$CDM halo displays a highly concentrated central core with a steeper projected density profile, whereas the IDE counterpart appears significantly more extended and diffuse, with a lower peak surface mass density. The surrounding substructures and nearby filamentary features also appear less sharply defined in the IDE simulation.

This behaviour is fully consistent with the results obtained for the matter power spectrum, where the IDE model shows enhanced clustering on large scales but a suppression of power in the quasi-linear and non-linear regimes. Physically, the reduced concentration of collapsed structures can be understood as a consequence of the energy transfer from dark matter to dark energy in the IDE model considered here ($\xi<0$). The interaction effectively decreases the amount of gravitating dark matter available at late times, reducing the depth of the gravitational potential wells and consequently lowering the efficiency of halo collapse and virialisation. As a result, haloes in the IDE cosmology tend to be less massive and less concentrated than their $\Lambda$CDM counterparts. Our results provide additional evidence that dark-sector interactions can significantly modify the internal properties of haloes and the non-linear distribution of matter, even when the large-scale cosmic web remains qualitatively similar to that of $\Lambda$CDM. A similar suppression of halo concentrations in other interacting or coupled dark-energy scalar-field cosmologies has previously been reported in~\cite{Baldi:2010pq,Zhao:2022ycr}.

Figure~\ref{fig:hmf} shows the halo mass function at $z=0$ for both the IDE and $\Lambda$CDM simulations, compared with the analytical fitting function of~\cite{Despali:2015yla}. The $\Lambda$CDM simulation is in very good agreement with their prediction over the full resolved mass range, with residuals typically at the $\sim5$--$10\%$ level over the central mass range and slightly larger scatter at the highest masses due to small-number statistics. This agreement validates both the numerical implementationion adopted in this work and the halo identification procedure used in the analysis.

In contrast, the IDE cosmology exhibits a clear suppression in the abundance of haloes relative to the $\Lambda$CDM case. The red curve shows the halo mass function directly computed from the halo finder using the particle mass at $z=0$. As shown in the lower panel, the ratio between the IDE simulation and the fitting function of~\cite{Despali:2015yla} remains approximately at the level of $\sim0.6$ over the mass interval $\sim 2\times10^{13}$--$10^{15}\,M_{\odot}h^{-1}$, corresponding to a deficit of roughly $\sim40\%$ in the halo number counts. The suppression becomes more pronounced towards the high-mass end, indicating that the formation of the most massive virialised structures is particularly sensitive to the IDE interaction.

This behaviour is a direct consequence of the energy transfer from dark matter to dark energy in the IDE model considered here ($\xi<0$). The interaction effectively reduces the amount of gravitating dark matter available at late times, weakening the growth of deep gravitational potential wells and decreasing the efficiency of halo collapse and hierarchical mass accretion. As a result, although the IDE cosmology predicts enhanced linear clustering on large scales, the non-linear formation of massive haloes is significantly suppressed relative to the standard $\Lambda$CDM scenario.

These results are fully consistent with the behaviour observed in the matter power spectrum and projected density maps, where the IDE model displayed suppressed clustering in highly non-linear regions and less concentrated halo structures. Overall, the halo mass function analysis demonstrates that interacting dark energy cosmologies can substantially modify the connection between linear growth and non-linear structure formation, reinforcing the importance of dedicated N-body simulations for obtaining reliable predictions beyond the linear regime.

We note that a direct comparison between our simulated halo mass function and observational cluster-count data is not straightforward, since observed cluster abundances rely on scaling relations between halo mass and survey observables, which are affected by significant systematic uncertainties. In particular, the mass-bias parameter, $(1-b)$, relating inferred and true halo masses, remains poorly constrained, with estimates ranging from $\sim0.6$ to $\sim0.9$ depending on the calibration method.

\section{Conclusions}
\label{final}

In this work, we performed gravity-only N-body simulations of the interacting dark energy model characterised by the coupling $Q=\xi {\cal H}\rho_x$, using a modified implementation of \texttt{RAMSES} designed to consistently incorporate the IDE background evolution and the effective dark matter density evolution. Our approach exploits the fact that, for the class of IDE models considered here, the dark matter Euler equation remains identical to that of $\Lambda$CDM and therefore does not generate an additional fifth force. We also verified that dark energy perturbations remain negligible on the scales relevant for the simulations, allowing the interaction to be effectively described through the modified expansion history and the time-dependent evolution of the dark matter particle mass.

The main motivation for this work arises from the current limitations in the modelling of non-linear structure formation in interacting dark energy cosmologies. Although IDE models have been extensively studied as possible mechanisms for alleviating the Hubble tension, their implications for non-linear observables and the $S_8$ tension remain considerably less understood. Current weak-lensing and large-scale-structure analyses typically remove highly non-linear scales or model them using prescriptions calibrated within $\Lambda$CDM, such as \texttt{HMcode} and standard halo-model fitting functions. Consequently, present constraints on IDE cosmologies in the non-linear regime are not fully model-consistent.

Our simulations show that the IDE interaction produces significant and scale-dependent modifications to non-linear structure formation. In the matter power spectrum, we find that the IDE cosmology enhances clustering on large and quasi-linear scales while suppressing power in the deeply non-linear regime. The comparison with the non-linear \texttt{HMcode} predictions further demonstrates that the mapping between linear and non-linear clustering is modified in a non-trivial way in IDE cosmologies. In particular, despite the substantially larger linear $\sigma_8$ value of the IDE model, the non-linear evolution partially suppresses the excess clustering at small scales.

The projected dark matter density maps reveal additional signatures of the interaction. While the large-scale filamentary structure remains qualitatively similar to that of $\Lambda$CDM, haloes in the IDE cosmology appear systematically less concentrated and more diffuse. This behaviour is fully consistent with the suppression observed in the non-linear matter power spectrum and can be physically interpreted as a consequence of the energy transfer from dark matter to dark energy, which reduces the efficiency of late-time halo collapse and virialisation.

The halo mass function analysis further supports this picture. Relative to both the $\Lambda$CDM simulation and the analytical fitting function calibrated in standard cosmology, the IDE simulation exhibits a substantial suppression in the abundance of massive haloes, reaching approximately $\sim40\%$ over the mass range explored in this work. This result demonstrates that the IDE interaction significantly modifies the non-linear collapse history and the formation of virialised structures, particularly at the high-mass end where halo growth is most sensitive to late-time gravitational evolution.

Overall, our results demonstrate that the impact of interacting dark energy on non-linear observables cannot be captured by a simple rescaling of the linear clustering amplitude or by standard $\Lambda$CDM-calibrated fitting prescriptions. The non-linear effects induced by the interaction alter the relation between linear growth, halo formation, and small-scale clustering in a scale-dependent manner. This suggests that cosmological constraints on IDE models derived from weak lensing, cluster counts, and galaxy clustering should be interpreted with caution whenever the non-linear regime is described using prescriptions calibrated exclusively within $\Lambda$CDM.

Our work provides a first step towards a fully model-consistent treatment of non-linear structure formation in interacting dark energy cosmologies. Future work will extend this analysis to larger simulation volumes, higher mass resolution, improved halo statistics, and direct predictions for weak-lensing observables and cosmological inference pipelines. This will allow a more robust assessment of whether the large values of $S_8$ often inferred in IDE cosmologies represent genuine physical predictions of the model or instead reflect limitations of the non-linear modelling currently adopted in observational analyses.
	
\begin{acknowledgments}
CvdB is supported by the Lancaster–Sheffield Consortium for Fundamental Physics under STFC grant: ST/X000621/1. EDV is supported by a Royal Society Dorothy Hodgkin Research Fellowship. This article is based upon work from the COST Action CA21136 - ``Addressing observational tensions in cosmology with systematics and fundamental physics (CosmoVerse)'', supported by COST - ``European Cooperation in Science and Technology''.  
RCN thanks the financial support from the Conselho Nacional de Desenvolvimento Científico e Tecnológico (CNPq, National Council for Scientific and Technological Development) under the project No. 304306/2022-3, and the Fundação de Amparo à Pesquisa do Estado do RS (FAPERGS, Research Support Foundation of the State of RS) for partial financial support under the project No. 23/2551-0000848-3. BL is supported by the ERC Advanced Grant `UNCA' (UKRI Frontiers Research Guarantee No.~EP/Z533877/1) and the UK STFC Consolidated Grant ST/X001075/1. We acknowledge IT Services at The University of Sheffield for the provision of services for High Performance Computing.
\end{acknowledgments}
	
	

	\bibliographystyle{apsrev4-1}
	\bibliography{biblio}
	
\end{document}